\documentclass[prb,twocolumn,showpacs,superscriptaddress,preprintnumbers,amssymb]{revtex4}
\usepackage{graphicx}
\usepackage{dcolumn}
\usepackage{bm}
\usepackage{wasysym}

\newcommand{\beq}{\begin{equation}}
\newcommand{\eeq}{\end{equation}}
\newcommand{\beqn}{\begin{eqnarray}}
\newcommand{\eeqn}{\end{eqnarray}}

\begin{document}
\title{Bond algebraic liquid phase in strongly correlated multiflavor cold atom systems}
\author{Cenke Xu}
\affiliation{Department of Physics, University of California,
Berkeley, CA 94720} \author{Matthew P. A. Fisher}
\affiliation{Kavli Institute of Theoretical Physics, University of
California, Santa Barbara, CA, 93106}
\date{\today}
\begin{abstract}
When cold atoms are trapped in a square or cubic optical lattice,
it should be possible to pump the atoms into excited $p-$level
orbitals within each well.  Following earlier work,  we explore
the metastable equilibrium that can be established before the
atoms decay into the $s-$wave orbital ground state. We will
discuss the situation with integer number of bosons on  every
site, and consider the strong correlation ``insulating" regime. By
employing a spin-wave analysis together with a new duality
transformation, we establish the existence and stability of a
novel gapless ``critical phase", which we refer to as a ``bond
algebraic liquid". The gapless nature of this phase is stabilized
due to the emergence of symmetries which lead to a quasi-one
dimensional behavior. Within the algebraic liquid phase, both bond
operators and particle flavor occupation number operators have
correlations which decay algebraically in space and time. Upon
varying parameters, the algebraic bond liquid can be unstable to
either a Mott insulator phase which spontaneously breaks lattice
symmetries, or a $\mathbb{Z}_2$ phase. The possibility of
detecting the algebraic liquid phase in cold atom experiments is
addressed. Although the momentum distribution function is
insufficient to distinguish the algebraic bond liquid from other
phases, the density correlation function can in principle be used
to detect this new phase of matter.

\end{abstract}
 \maketitle

\section{introduction}

Interacting bosonic systems hopping on two or three dimensional
lattices ($d=2,3$) typically have one of two ground states:
superfluid or a Mott insulator \cite{fisher1989}. Searching for
bosonic phases other than these two has attracted much attention,
especially gapless bosonic phases which do not break the global
$U(1)$ symmetry. In such gapless featureless bosonic phases, the
correlation functions between physical operators usually fall off
algebraically. Being the lattice boson analog to the algebraic
spin liquids which can occur in spin models, these phases can be
referred to as algebraic boson liquids.  So far several algebraic
liquid phases have been realized in toy models. In $d=3$ stable
algebraic spin liquids with gapless photon excitations have been
realized in dimer model on a cubic lattice \cite{sondhi2003}, as
well as spin models on a pyrochlore lattice
\cite{wen2003,hermele2004}.  In two-dimensional  bosonic models
with strong ring exchange terms, an ``excitionic bose liquid"
phase has been proposed in\cite{paramekanti,balents2005c}.  More
recently another type of stable boson algebraic liquid phase with
softened graviton excitations has been shown to be stable on an
fcc lattice \cite{xu2006}.  Establishing the stability of such
algebraic liquid phases is oftentimes quite subtle, since the
gapless excitations are not due to the breaking of a continuous
symmetry - they are not Goldstone modes \cite{goldstone}.

The proliferation of topological defects in $d>1$ often drives the
instability of algebraic spin liquids.   For example, as pointed
out by Polyakov \cite{polyakov1977,polyakov1987}, the monopole
(instanton) excitation in compact electrodynamics in $d=2$ will
always proliferate and will gap out the photon excitation.  On the
other hand, if a bosonic model has a set of emergent symmetries at
low energies which leads to a quasi-one dimensional behavior, as
in the excitonic Bose liquid studied in
\cite{paramekanti,balents2005c}, stability can be achieved by
analogy with Luttinger liquids in $d=1$. In the present paper we
explore a new algebraic boson liquid phase in which the
gaplessness is protected by a similar mechanism of an emergent
quasi one-dimensional behavior.


The model under consideration is motivated by cold bosonic atoms
systems which can be trapped in square or cubic lattices formed by
laser beams. The single particle ground state wavefunction within
each well of the optical lattice will have an approximate $s-$wave
symmetry, with wave function $\phi_0(r)\sim\exp(-\alpha r^2)$. The
first excited states are three fold degenerate $p-$wave states,
with wave functions $\phi_x(r)\sim x\exp(-\alpha r^2)$,
$\phi_y(r)\sim y\exp(-\alpha r^2)$ and $\phi_z(r)\sim
z\exp(-\alpha r^2)$. The wave function $\phi_x$ extends further in
the $x$ than in the $y$ and $z$ directions, and so will
preferentially hop to adjacent wells along the $x-$axis. As
discussed in \cite{girvin2005}, although the $p-$wave states are
not the ground states, the life time for these states can be much
longer than the average tunnelling time between neighboring
optical lattice sites. If all of the atoms are pumped from the
ground state to the $p-$wave states, they will establish a
metastable equilibrium which can survive for considerable time.
Here, following \cite{girvin2005}, we explore the properties of
bosonic atoms trapped in such excited $p-$wave states.

In previous studies, several phases have been identified for this
system, including a superfluid phase with 1 dimensional
$\mathbb{Z}_2$ gauge symmetry \cite{girvin2005}, the Mott
insulator phase,  a novel stripe phase \cite{wu2006}. A supersolid
phase has been predicted in even higher excited bands in cold atom
systems in optical lattices \cite{scarola2005}. In our work, the
same system with strong interaction is revisited. The interaction
induces a big gap to the excitations which change the total number
of bosons on each site, and hence precludes superfluidity. We
establish that in addition to the Mott insulating state, a novel
bond-algebraic liquid phase can be realized in both $d = 2$ and $d
= 3$ systems. In these phases, bond-bond correlation falls off
algebraically, but the correlation function is very anisotropic.
We examine the instabilities of this gapless bond liquid phase
towards both Mott Insulator and the $\mathbb{Z}_2$ phase.  When
the average filling per site is not a multiple of the spatial
dimensions $d$, we find that, over a range of parameters, the bond
liquid phase is stable. In $d = 2$, this liquid phase has a nice
self-dual structure, which closely resembles the exciton bose
liquid phase studied earlier \cite{paramekanti}.   In the $d = 3$
case, we obtain a dual representation in terms of vortex loops
which can hop on the dual lattice.

This paper is organized as follows.  In section II, the low energy
effective Hamiltonian is derived, and the quasilocal symmetry of
the low energy Hamiltonian is discussed. In section III, we
implement a spin-wave analysis which enables us to access and
explore the bond algebraic liquid phase.  In section IV, dual
representations are derived in both $d = 2$ and $d = 3$. Section V
is devoted to an analysis of the stability of the bond liquid
phase. Instabilities towards both the $\mathbb{Z}_2$ phase and the
Mott Insulator are also discussed. Employing the dual
representation we identify the crystalline symmetry breaking
pattern of the Mott insulator. In section VI, we consider
modifying the model by doping with $s-$wave particles and
softening the constraint on the particle number fluctuations
within each well. Several additional phases are thereby obtained.
Finally, section VII is devoted to a discussion of the prospects
for experimental detection of the algebraic bond liquid phase.

\section{Models and Symmetries}

The Hamiltonian describing the cold atom system includes the
kinetic term, and the $s-$wave scattering: \beqn H = \int d^3x
\psi^\dagger(x)(- \frac{\hbar^2}{2m}\nabla^2 - \mu + V_T(x))
\psi(x) \cr\cr + \frac{1}{2}\frac{4\pi a_s\hbar^2}{m}\int
d^3x\psi^\dagger(x)\psi^\dagger(x)\psi(x)\psi(x) ,
\label{atomhamil}\eeqn where $\mu$ is the chemical potential which
determines the average filling of atoms $\bar{n}$, $V_T(x)$ is the
optical trap potential, and $a_s$ is the $s-$wave scattering
length. We now assume that within each well all of the atoms are
maintained in the $p-$wave excited states. One can then expand the
atom field operators in terms of the atomic $p-$wave states on
each site \beqn \psi(x) = \sum_{i}\sum_{a} (-1)^{\sum_bi_b}
d^{(a)}_i\phi_a(x - i), \eeqn
with $a,b = x,y$ in $d = 2$ system,
and $a,b = x,y,z$  in $d = 3$ system.
Here $d^{(a)}_i$ denotes a boson destruction operator of flavor
$a$ on site $i$, and $i_x,i_y,i_z$ denote the $x,y,z$ coordinates of the lattice site $i$.  Inserting this expansion
into equation (\ref{atomhamil}), the kinetic term generates an
on-site energy term for each flavor of atoms: \beqn H_o =
\sum_{i,a}\epsilon_{a}d^{(a)\dagger}_id^{(a)}_i,\cr\cr \epsilon_a
= \int d^3x \phi_a(x) (- \frac{\hbar^2\nabla^2}{2m} + V_T(x))
\phi_a(x).\label{onsiteenergy} \eeqn as well as a (dominant)
nearest neighbor hopping term of the form, \beqn H_h = -
\sum_i\sum_{a,b,c} t^\prime_{a,b,c}
(d^{(a)\dagger}_id^{(b)}_{i+\hat{c}}+h.c.),\cr\cr t^\prime_{a,b,c}
= \int d^3x \phi_a(x) (- \frac{\hbar^2\nabla^2}{2m} + V_T(x))
\phi_b(x - \hat{c}). \label{atomhop}\eeqn

The on-site energy for each flavor of atom needs not be the same,
if the lattice symmetry is broken. For instance, if the laser
beams in $x$ and $y$ directions have different wavelengths or
amplitudes, $p_x$ particles should have a different energy from
the $p_y$ particles. The hopping in (\ref{atomhop}) is
preferentially along one of the three axes, that is the integral
in (\ref{atomhop}) is small unless $a = b = c$. The kinetic term
is dominated by the very anisotropic hopping terms, for instance
$p_x$ particles can only hop easily in the $\hat{x}$ direction.
Since the other hopping terms will be small, we expect that over a
large range of energies they will be unimportant. So we henceforth
set these other hopping terms to zero. This is the central
assumption of all that follows. Thus the hopping term is
approximated as \beqn H_h = -t^\prime \sum_i\sum_{a}
(d^{(a)\dagger}_id^{(a)}_{i+\hat{a}}+h.c.). \eeqn Due to the form
of the $p-$wave orbitals the remaining single hopping strength,
$t^\prime$ should be positive.

In addition to the kinetic energy, the ($s-$wave scattering)
interaction between the atoms will generate an on-site repulsion,
as well as an intra well pair-tunnelling term between different
flavors, \beqn H_{s} = \frac{1}{2}\frac{4\pi a_s\hbar^2}{m}\int
d^3x\psi^\dagger(x)\psi^\dagger(x)\psi(x)\psi(x)\cr\cr \sim
\sum_i\sum_{a,b}U_{ab}n^{(a)}_in^{(b)}_i + \sum_{a\neq b}\gamma
(d_i^{(a)\dagger}d_i^{(a)\dagger}d_i^{(b)}d_i^{(b)} + h.c.)
\label{swave}\cr \eeqn  The $U_{ab}$ term encodes the repulsion
between particles with the same flavor as well as between
different flavors. The $\gamma$ term is the flavor mixing term,
where, for example, two $p_x$ particles can be converted into two
$p_y$ particles, and vice versa. The flavor mixing is due to the
fact that, despite the vanishing overlap between the wave
functions $\phi_x(r)$ and $\phi_y(r)$, the overlap integral $\int
d^3x (\phi_x(r)^\ast)^2(\phi_y(r))^2$ does not vanish. Angular
momentum conservation does not rule out this pair conversion term
because, while the $p_x$ state is an eigenstate of $L_x$,  $p_y$
is not. Moreover, single particle flavor conversion is strictly
forbidden being odd under parity symmetry, \beqn P_x\phi_x
\rightarrow - \phi_x, P_x \phi_y \rightarrow \phi_y, \cr P_y\phi_y
\rightarrow - \phi_y, P_y \phi_x \rightarrow \phi_x.\eeqn Because
the integral $\int d^3x (\phi_x(r)^\ast)^2(\phi_y(r))^2$ is
positive, $\gamma$ will be positive.

The interflavor and the intraflavor interactions will generally
not be the same. For simplicity, we assume that the interaction
matrix $U_{ab}$ together with the chemical potential can be
adequately approximated by a term of the form, \beqn U(\sum_a
n^{(a)}_i - \bar{n})^2 + \sum_a u(n^{(a)}_i - \bar{n}_a)^2,\cr\cr
\sum_a\bar{n}_a = \bar{n}.\label{u}\eeqn Here $\bar{n}_a$ is the
average filling of each flavor. If all flavors are symmetric,
$\bar{n}_a = \bar{n}/d$. Also, let us not forget that since the
wave functions of $p_a$ particles have comparatively large overlap
between two nearest neighbor sites in the same $\hat{a}$ axis ($a
= x$, $y$, $z$), the $s$-wave scattering can generate a small
off-site repulsion interaction between $p_a$ particles along
$\hat{a}$ direction $\delta H = \sum_{i,a}\delta u n^{(a)}_i
n^{(a)}_{i+\hat{a}}$, this anisotropic nearest neighbor repulsion
is supposed to be much stronger than the nearest neighbor
repulsion between $s$-wave particles, which is usually negligible.

Due to the flavor mixing $\gamma$ term in (\ref{swave}), the boson
number $n_x - n_y$ is not conserved. However, as we will show
below, in the phase of primary interest, the algebraic bond liquid
phase, the flavor mixing term can be irrelevant, therefore in this
phase the boson numbers of each flavor of bosons can be considered
as independent conserved quantities, and each flavor of boson has
its own separate average filling.

Let us assume $U$ is the biggest energy scale in (\ref{u}), and
$u$ is a small energy penalty for density fluctuation around
average filling of each flavor, then the main effect of the term
in (\ref{u}) is to keep the number of atoms in each well close to
the average filling $\bar{n}$, which we will always take as an
integer. Moreover, we consider the strong correlation limit, where
$U$ is much larger than the kinetic energy term, $t^{\prime}$.
Since the average filling for each flavor is $\bar{n}_a$, when
$\bar{n}_a$ is not an integer, the system is effectively
``fractionally" filled. Below, we will first focus on the case
with $\bar{n} \geq 2$, deferring a discussion of the behavior in
the case with $\bar{n} = 1$, which is special because the
intrawell pair conversion term alone cannot directly operate on
the low energy subspace.

The full Hubbard type Hamiltonian without $\delta H$ then takes
the form, \beqn H = H_h + H_U + H_s = \cr\cr - t^\prime
\sum_i\sum_{a} (d^{(a)\dagger}_id^{(a)}_{i+\hat{a}}+h.c.) +
U(\sum_a n^{(a)}_i - \bar{n})^2 \cr\cr + \sum_a u(n^{(a)}_i -
\bar{n}_a)^2 + \sum_i \sum_{a\neq b}\gamma
((d^{(a)\dagger}_i)^2(d^{(b)}_i)^2 + h.c.) . \label{atomfull}\eeqn
In the large $U=\infty$ limit, there will be a hard constraint on
the Hilbert space, with each site occupied by exactly $\bar{n}$
particles if $\bar{n}$ is integer.  For large but finite $U$ we
will project the Hubbard Hamiltonian into this constrained Hilbert
space. A single particle hopping event will take one out of this
low energy Hilbert space. At the second order, a ``trivial"
process which hops a $p_a$ particle at site $i$ to site
$i+\hat{a}$, and then hops back to $i$ can take place. This
hopping process does not change the distribution of particles, but
two $p_a$ particles at two nearest neighbor sites along $\hat{a}$
direction can benefit from this ``trivial" process. Therefore this
hopping generates a small effective attractive interaction between
two particles $\delta H^\prime = -  \sum_{i,a} \delta v n^{(a)}_i
n^{(a)}_{i+\hat{a}}$, $\delta v \sim t^{\prime 2}/U$. This
off-site attractive interaction competes against $\delta H$, and
generally drives the system into orbitally ordered phase. if
$\delta H$ is dominated by $\delta H^\prime$, ferromagnetic
distribution of flavors is favored; otherwise if $\delta H^\prime$
wins, $p_a$ particle at site $i$ does not want to see the same
flavor of particle stay at site $i + \hat{a}$, in 2 dimensional
case the system will be driven to the antiferromagnetic order of
$n_x - n_y$.

\begin{figure}
\includegraphics[width=2.0 in]{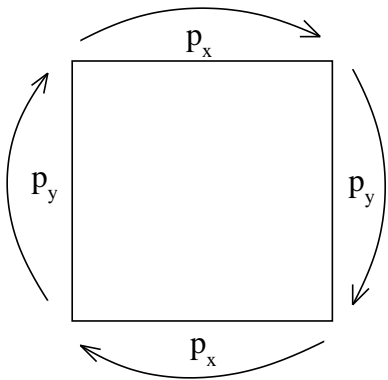}
\caption{A pictorial representation of the ring exchange term in
(\ref{ring}). If a $p_x$ particle hops to the right, another $p_x$
particle must hop back to left, and similarly for the $p_y$
particles. \label{ringfig}}
\end{figure}

From now on let us focus on the case with $\delta u \approx \delta
v$, thus there is no obvious tendency to orbital order, the
physics is controlled by higher order perturbations (our main
result does not require $\delta u$ exactly equals to $\delta v$,
we will argue that a small residual off-site density interaction
does not destabilize the phase of main interest). At forth order
perturbation, a flavor-mixing ring exchange term of the form,
\beqn H_{ring} = \cr\cr -\tilde{t} \sum_{i}\sum_{a\neq b}
(d^{(a)}_id^{(a)\dagger}_{i+\hat{a}}d^{(a)}_{i +
\hat{a}+\hat{b}}d^{(a)\dagger}_{i+\hat{b}} d^{(b)}_{i+ \hat{b}}
d^{(b)\dagger}_i d^{(b)}_{i+\hat{a}}d^{(b)\dagger}_{i +
\hat{a}+\hat{b}} + h.c.) \cr \label{ringd} \eeqn will survive the
low energy projection. Here, $\tilde{t} \sim (t^\prime)^4/U^3$.
The ring exchange in the $XY$ plane is depicted in Fig.
\ref{ringfig}. This term looks fairly complicated, but if we make
the usual approximation, replacing the lattice boson operators
with rotors $d^{(a)}\sim e^{i\theta_a}$, it will appear much
simpler.   The effective rotor Hamiltonian reads, \beqn H_{r} =-t
\sum_i\sum_{a\neq b}\cos(\partial_a\partial_b(\theta_{ai} -
\theta_{bi})) + \sum_{a}u(n_{ai} - \bar{n}_a)^2 \cr + \sum_{a\neq
b}\gamma\cos(2(\theta_{ai} - \theta_{bi})) . \label{ring}\eeqn All
the derivatives in (\ref{ring}) should be understood as lattice
derivative, i.e. $\partial_{\hat{a}}\theta(x) = \theta(x +
\hat{a}) - \theta(x)$.

With $\bar{n} \ge 2$, the Hamiltonian $H_r$ in equation (\ref{ring})
is the starting point of our subsequent analysis. We first
consider the symmetries of this Hamiltonian, initially tuning
$\gamma$ to be zero (the legitimacy of doing this will be
discussed later).  The symmetry in (\ref{ring}) can then be
summarized as follows. When $d = 2$, the Hamiltonian is invariant
under the transformation, \beqn \theta_x \rightarrow \theta_x +
f_1(y) + g_1(x) + \alpha(x,y)\cr \theta_y \rightarrow \theta_y +
f_2(x) + g_2(y) + \alpha(x,y)\label{symmetry2}, \eeqn and for $d =
3$ an analogous form, \beqn \theta_x \rightarrow \theta_x +
f_1(y,z) + g_1(x) + \alpha(x,y,z)\cr \theta_y \rightarrow \theta_y
+ f_2(x,z) + g_2(y) + \alpha(x,y,z)\cr \theta_z \rightarrow
\theta_z + f_3(x,y) + g_3(z) + \alpha(x,y,z) .
\label{symmetry}\eeqn Here $f_i,g_i$ and $\alpha$ are arbitrary
functions of their arguments. We can understand these symmetries
as follows: the $f_i$ are functions of  $d-1$ coordinates
reflecting the fact that each flavor can only hop in one
dimension.  The conservation laws associated with these $U(1)$
symmetries correspond, for example, to the conserved total number
of $x-$flavor particles at each fixed value of $y$ and $z$. By
contrast, the symmetries associated with the
 $g_i$ functions are additional emergent symmetries, present only after projecting into
 the low energy Hilbert space with a fixed number of particles on each site.
Physically, in the projected Hilbert space each flavor of boson
cannot hop freely: When one $p_x$ particle hops in the $+\hat{x}$
direction, another $p_x$ particle must hop in the opposite
direction (see Fig 1). Finally, the $\alpha$ functions, depending
on all $d$ of the spatial coordinates, are (unphysical) gauge
symmetries corresponding to the hard projection imposed on the
system, $\sum_a n_a = \bar{n}$.

If the $\gamma$ term is present, the continuous symmetries in
(\ref{symmetry2}) and (\ref{symmetry}) are broken down to
$\mathbb{Z}_2$ symmetries.  All the functions $f_i$ and $g_i$ can
only take two values, $0$ or $\pi$.

\section{Algebraic phases}

We first analyze the Hamiltonian in (\ref{ring}) in the special
case with $\gamma = 0$.  As we shall see, this enables us to
access a novel algebraic bose liquid phase.  We will then turn to
the effects of a small but non-zero $\gamma$, and find that under
certain conditions $\gamma$ is an irrelevant perturbation. Under a
coarse graining in space and time it will scale to zero.

To this end, we implement a ``spin-wave" approximation, expanding
the cosine terms in (\ref{ring}).  The resulting action describes
a Gaussian theory, with a Euclidian Lagrangian of the form, \beqn
L_{d = 2} = \sum_{a = x}^y\frac{K}{2}(\partial_\tau\theta_a)^2 -
\frac{K}{2}(\partial_x\partial_y(\theta_x - \theta_y))^2,\cr\cr\cr
L_{d = 3} = \sum_{a = x}^z\frac{K}{2}(\partial_\tau\theta_a)^2 -
\sum_{a\neq b} \frac{K}{2}(\partial_a\partial_b(\theta_a -
\theta_b))^2. \label{gaussian} \eeqn Here we have rescaled space
and time to set the velocity to unity. There is then just one
remaining dimensionless parameter, $K$. Such a quadratic
Lagrangian in which all terms involve derivatives of the fields,
describes a scale invariant phase at long lengthscales. In a sense
it can be viewed as a ``fixed point" Lagrangian. The resulting
phase is the desired algebraic boson liquid.

The gapless modes of this Guassian theory can be readily
calculated. In $d = 2$ there are two modes at each momenta with
frequencies given by, \beqn \omega_0^2 = 0, \cr\cr \omega_1^2 \sim
\sin(k_x/2)^2\sin(k_y/2)^2.\eeqn The first nondispersing mode
$\omega_0$ is an unphysical gauge mode, corresponding to the
function $\alpha(x, y)$ in (\ref{symmetry2}).  The second mode
vanishes along both coordinate axes in momentum space, as required
by the symmetries encoded in the functions $f_i$ and $g_i$ in
(\ref{symmetry2}).   This mode  gives rise to a quasi one
dimensional behavior, which is ultimately responsible for the
stability of the novel bond algebraic liquid phase. If $\omega_1$
remains gapless in the presence of all allowed perturbations, this
algebraic liquid phase is stable. The double intraflavor
conversion term and a set of ``vertex" operators present in a dual
representation (see below) can potentially destabilize the
algebraic phase and gap out the mode $\omega_1$. The stability of
the algebraic liquid phase will be discussed later.

In $d = 3$, in addition to the unphysical mode, $\omega_0 = 0$,
two other modes are obtained at each momenta. At small momentum
$k$ the dispersion relations take the form,
 \begin{eqnarray} \omega_1^2
\sim  2(k_x^2k_y^2 + k_x^2k_z^2 + k_y^2k_z^2), \cr\cr \omega_2^2
\sim \frac{3}{2}k_x^2k_y^2k_z^2\frac{k^2}{k_x^2k_y^2 + k_x^2k_z^2
+ k_y^2k_z^2}.\cr
\end{eqnarray}
Notice that $\omega_1$ vanishes along every coordinate axis in
momentum space, whereas $\omega_2$ vanishes on every coordinate
plane.  It is $\omega_2$ that encodes the quasi-1d behavior, and
as in $d=2$ will be the ultimate reason for the stability of the
algebraic liquid phase.

Due to the symmetries in (\ref{symmetry2}) and (\ref{symmetry}),
many correlation functions vanish at this Gaussian fixed point.
For instance, the correlation function between boson operators
$d^{(x)\dagger}$ and $d^{(x)}$ vanishes, as it breaks these
symmetries. For the same reason, the correlation functions between
two pair conversion operators vanish.   But this does not imply
that this operator is irrelevant, because two adjacent pair
conversion operators can generate potentially relevant operators,
which will be discussed later.

\begin{figure}
\includegraphics[width=2.5 in]{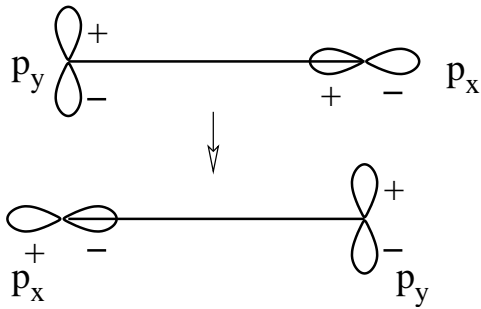}
\caption{A pictorial representation of the bond operator $B_{x}$.
If on one site there is one $p_x$ particle and on its adjacent
site there is one $p_y$ particle, $B_{x}$ will exchange the $p_x$
and $p_y$ particles on the two sites. \label{002dside}}
\end{figure}

The only correlators which do not vanish are correlation functions
between bond operators. In the $d = 2$ case, the bond operators
are defined as $B_{x} = \cos(\partial_x(\theta_x - \theta_y))$ and
$B_y = \cos(\partial_y(\theta_x - \theta_y))$. In the $d = 3 $
case, the bond operators are defined in a similar way, for
instance, $B_{xy,x} = \cos(\partial_x(\theta_x - \theta_y))$. If
on one site there is one $p_x$ particle and on its adjacent site
there is one $p_y$ particle, $B_{x}$ will exchange the $p_x$ and
$p_y$ particles on the two sites (Fig. \ref{002dside}).

The correlation functions between bond operators are \beqn d = 2:
\langle B_{x}(0,0,0)B_{x}(\tau,0,y)\rangle \sim \frac{1}{|\tau^2 +
y^2|^{1/(\pi^2K)}},\cr\cr d = 3: \langle
B_{xy,x}(0,0,0,0)B_{xy,x}(\tau,0,y,0)\rangle \cr\cr\sim
\frac{1}{|y^2 + \tau^2|^{\eta^\prime/K}}. \eeqn Here, the
dimensionless number $\eta^\prime$ depends on the form of the
fixed point action on the lattice scale, that is not just the
small momentum form . Notice that the correlation functions
between two $B_{x}$ ($B_{xy,x}$) operators are only nonzero when
they are at the same $(\tau, y)$ plane.

\section{dual representation}

\subsection{d = 2}

In the previous section the Gaussian fixed point was obtained by
expanding the cosine terms and allowing the fields to take on any
real value. Here we address the legitimacy of this procedure.
Specifically, we must take into account that the $\theta$ fields
are really phase angles (ie. compactified bosons) with the
identification $\theta = \theta + 2\pi \mathbb{Z}$.   This implies
that particular topological defects will be allowed.  The effect
of such defects can be most conveniently addressed by passing to a
dual representation, wherein one can identify ``vertex" operators
which insert such topological defects. The dual transformation of
the theory should solve the constraint $\sum_{a} n^{(a)}_i =
\bar{n}$.  In $d = 2$ we define dual variables $\phi$ and $N$ as
follows, \beqn n_x - \bar{n}_x =
\partial_x\partial_y\phi, \cr n_y - \bar{n}_y = -
\partial_x\partial_y\phi,\cr
\partial_x\partial_y(\theta_x - \theta_y) = N, \eeqn
so that the constraint is automatically satisfied. Here, $\phi$
and $N$ are both defined on the dual site (plaquette centers). We
can check the commutator and see that $N$ and $\phi$ are a pair of
conjugate variables.  As defined $\phi$ takes on only discrete values,
whereas $N$ lives on the interval $0$ to $2\pi$.  But we will
henceforth allow $\phi$ (and $N$) to roam over all the reals,
letting $\phi \rightarrow \varphi$ and
approximately imposing the discreteness by the addition
of terms in the action.

In terms of such a  ``coarse grained" field, $\varphi$, the Gaussian part of the dual action
reads \beqn L_{d = 2} = \frac{1}{2K}(\partial_\tau\varphi)^2 +
\frac{1}{2K}(\partial_x\partial_y\varphi)^2 .
\label{dualgaussian}\eeqn
When $\bar{n}$ is integer, and both $\bar{n}_x$
and $\bar{n}_y$ are integers, the $\phi_i$ should all be integers.
Therefore in the dual Hamiltonian there is a vertex operator
$\cos(2\pi\varphi)$ on each site. However, since the dual Gaussian
action (\ref{dualgaussian}) has a similar symmetry as in
(\ref{symmetry2}), \beqn \varphi \rightarrow \varphi + f(x) +
g(y),\label{dualsymmetry}\eeqn the correlation function between
two such $\cos(2\pi\varphi)$ operators on different sites is zero.
The vertex operators which show up in the low energy theory are
dipole operators $V_{x} = \cos(2\pi\partial_x\varphi)$ and $V_y =
\cos(2\pi\partial_y\varphi)$. At the Gaussian fixed point, the
correlation function between two dipole operators is \beqn
V_{x}(0,0,0)V_{x}(\tau,0,y)\rangle \sim \frac{1}{|\tau^2 +
y^2|^{4K}}. \eeqn

When $\bar{n}_x$ and $\bar{n}_y$ are not integers, $\phi_i$ cannot
all be integers. For simplicity, let us assume $\bar{n}$ is even
and, $\bar{n}_x = \bar{n}/2 + 1/m$ and $\bar{n}_y = \bar{n}/2 -
1/m$, with $m$ taking on all the integers. The arrangement of
$\phi$ can be taken as $\phi = \mathbb{Z} + xy/m$. Now the
potentially relevant vertex operator in the dual action is
$\cos(2\pi\partial_x\varphi + \mathcal{B}_x)$. Here $\mathcal{B}_x$
is a Berry's phase, $\mathcal{B}_x = 2\pi y/m$. When $m > 1$, this
Berry's phase gives rise to an oscillating sign on the lattice,
hence the vertex operator $\cos(2\pi\partial_x\varphi +
\mathcal{B}_x)$ does not appear in the low energy coarse grained
theory. However, vertex operator $V_{qp,x} =
\cos(2q\pi\partial_{px}\varphi) \equiv \cos(2q\pi\varphi(p,0) -
2q\pi\varphi(0,0))$ (with integer $p,q$), as well as ``parallel jump" ($PL$)term
$V_{pl} = \cos(2\pi\partial_x\varphi(0,0) -
2\pi\partial_x\varphi(1,0))$ do not contain any oscillating sign
on the lattice as long as $qp$ is a multiple of $m$, and hence
should appear in the low energy theory.

The leading (most relevant) operators are $V_{1m,x}$ and
$V_{m1,x}$, as well as $V_{pl}$ operators. The correlation
functions between $V_{qp,x}$ and $V_{pl}$ are \beqn\langle
V_{qp,x}(0,0,0)V_{qp,x}(\tau,0,y)\rangle \sim \frac{1}{|\tau^2 +
y^2|^{q^2K\delta(p)}}\cr\cr \langle
V_{pl}(0,0,0)V_{pl}(\tau,0,y)\rangle \sim\frac{1}{|\tau^2 +
y^2|^{32K/3}}. \label{corre2d}\eeqn
Here $\delta(p)$ is an integral
which depends on $p$, \beqn \delta(p) = \frac{1}{2}\int_0^\pi dk_x
\frac{(2\sin(pk_x/2))^2}{\sin(k_x/2)}.\eeqn When $p$ is large,
$\delta(p)$ roughly scales as $\delta(p) \sim \ln p$, therefore,
in the case of large $p$, $V_{pl}$ is the most relevant operator
in the dual formalism.

\subsection{d = 3}

In the $d = 3$ case, the compactification of the phase angles can
also be dealt with in a dual representation. In the case with $d =
3$, we only focus on the situation with $\bar{n}_x = \bar{n}_y =
\bar{n}_z = \bar{n}/3$, as in this case, so long as $\bar{n} \neq
3\mathbb{Z}$, it is enough to guarantee a stable liquid phase. Let
us define dual variables $N_i$ and $\phi_i$ as

\begin{eqnarray} N_z =
\partial_x\partial_y(\theta_x - \theta_y),\cr N_y =
\partial_z\partial_x(\theta_z - \theta_x),\cr N_x = \partial_y\partial_z(\theta_y -
\theta_z),\cr\cr \partial_z\partial_x\phi_y -
\partial_z\partial_y\phi_x  = \bar{n}/3 - n_z,\cr \partial_x\partial_y\phi_z -
\partial_x\partial_z\phi_y = \bar{n}/3 - n_x,\cr \partial_y\partial_z\phi_x -
\partial_y\partial_x\phi_z = \bar{n}/3 - n_y,\cr\cr
[\phi_i, N_j] = i\delta_{ij} .\label{dual}
\end{eqnarray}

\begin{figure}
\includegraphics[width=2.5 in]{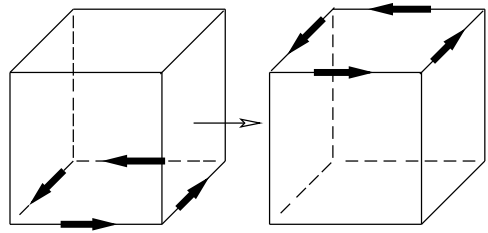}
\caption{The dual theory defined in (\ref{dual}) describes vortex
loops which can only hop across a cube, as shown above.
\label{loophop}}
\end{figure}

Here $N_x$, $N_y$ and $N_z$ can be interpreted as elements of
``vortex" lines in the $\theta_a$ configurations, along the
$\hat{x}$, $\hat{y}$ and $\hat{z}$ directions, respectively.
Moreover,  $\cos(\partial_z\partial_x\phi_y -
\partial_z\partial_y\phi_x)$ is a term which hops a unit size vortex
loop in the $xy-$plane along the $\hat{z}$ direction (Fig.
\ref{loophop}). Therefore the dual picture describes a system of
vortex-like loops which can hop in one dimension.  This is
analogous to the roton hopping model studied in $d=2$
\cite{balents2005}.

Various phases can be understood in terms of vortex loops in boson
systems. The normal three dimensional boson theory is dual to a
theory of vortex loops minimally coupled to a rank-2 antisymmetric
gauge field. The superfluid phase is a phase in which the vortex
loops tend to shrink, and the gapless rank-2 antisymmetric gauge
field corresponds to the Goldstone mode in the original theory. In
the 3 dimensional crystal (or Mott insulator) phase of the bosons,
the vortex loops proliferate and ``condense",  gapping out the
rank-2 antisymmetric gauge field through the Higgs mechanism. If
there are two flavors of vortex loops, a gapless boson liquid
phase can be realized, where the fluctuation of vortex loops
become the photon excitations in the photon boson liquid
\cite{senthil2005}. Here we show that when the vortex loops hop in
a perpendicular direction only, we can recover a dual
representation of the algebraic boson liquid of the original
$p-$wave boson system.

The Gaussian part of the dual vortex loop hopping action reads:

\begin{eqnarray}
L_{loop} = \cr\cr \sum_a\frac{1}{2K}(\partial_\tau\varphi_a)^2 +
\sum_{a \neq b, b\neq c, c\neq
a}\frac{1}{2K}(\partial_a(\partial_b\varphi_c -
\partial_c\varphi_b))^2  . \label{loop}\cr
\end{eqnarray}
Again,
$\varphi_a$ is a coarse grained field of $\phi_a$.
The symmetries of this dual Gaussian action are \beqn \varphi_x
\rightarrow \varphi_x + f_1(x,y) + f_2(x,z) +
\partial_x\alpha(x,y,z) ,\cr \varphi_y
\rightarrow \varphi_y + g_1(x,y) + g_2(y,z) +
\partial_y\alpha(x,y,z) , \cr \varphi_z \rightarrow \varphi_z + h_1(x,z)
+ h_2(y,z) +
\partial_z\alpha(x,y,z) .\label{symmetrydual} \eeqn

\begin{figure}
\includegraphics[width=1.8 in]{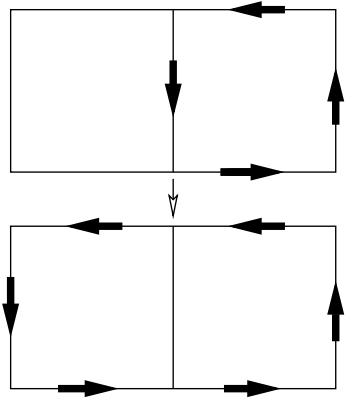}
\caption{The loop growing vertex operator can enlarge a vortex
loop. The arrow is the direction of the vortex.\label{loopgrow}}
\end{figure}

\begin{figure}
\includegraphics[width=1.8 in]{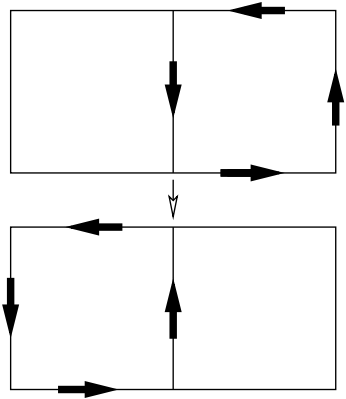}
\caption{The parallel jump vertex operator can jump a unit vortex
loop to its adjacent unit square. The arrow shows the direction of
the vortex.\label{loopside}}
\end{figure}

\begin{figure}
\includegraphics[width=2.3 in]{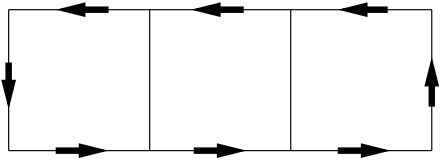}
\caption{The ``large" loop creation operator $G_{13,ab}$, which
creates a ``large" vortex loop covering three unit squares.
\label{loopbig}}
\end{figure}

Due to its definition in (\ref{dual}), $\phi_a$ only takes on
discrete values. Thus, as in the $d=2$ case, vertex operators are
allowed in the dual theory. Most of these vertex operators are
irrelevant at the Gaussian fixed point since the correlation
functions between them break the symmetries listed in
(\ref{symmetrydual}).  One important vertex operator with nonzero
correlations is a loop growing operator $G_{ab} =
\cos(2\pi(\partial_a\varphi_b -
\partial_b\varphi_a) + \mathcal{B})$, where $\mathcal{B}$ is again a  Berry's phase.
This loop growing term can enlarge, shrink, annihilate and create
a vortex loop (Fig. \ref{loopgrow}).  A Berry's phase term will
generally be present in the dual vertex operators, except when
$\bar{n} = 3\mathbb{Z}$ where it vanishes.   Since the filling of
the boson is $\bar{n}/3$ for each flavor, when $\bar{n}$ is not an
multiple of 3, the vertex operators which survive in the low
energy field theory should be  $G_{31,ab} =
\cos(6\pi(\partial_a\varphi_b -
\partial_b\varphi_a))$ and $G_{13,ab} = \prod_{i = 0}^{2}\cos(2\pi(\partial_{a}\varphi_b -
\partial_b\varphi_a)_{(a,b+i)})$. $G_{13,ab}$ is an operator that
creates a ``large" vortex loop which covers three unit squares
(Fig. \ref{loopbig}). Another potentially relevant vertex operator
is the parallel jumping term, denoted as $V_{pl}$. This term hops
a unit vortex loop parallel to its unit square (Fig
\ref{loopside}). Both the loop growing term and the $V_{pl}$ term
can only correlate in the direction perpendicular to the loop. The
correlation function for these two operators are,

\begin{eqnarray}
\langle G_{31,xy}(0,0,0,0)G_{31,xy}(\tau,0,0,z) \rangle \cr\cr
\sim \frac{1}{|\tau^2 + z^2|^{K\eta_1}} ,\cr\cr\cr \langle
G_{13,xy}(0,0,0,0)G_{13,xy}(\tau,0,0,z) \rangle \cr\cr \sim
\frac{1}{|\tau^2 + z^2|^{K\eta_2}}, \cr\cr\cr \langle
V_{pl}(0,0,0,0)V_{pl}(\tau,0,0,z) \rangle \sim \frac{1}{|\tau^2 +
z^2|^{K\eta_3}}\label{corre3d}.\cr
\end{eqnarray}

Again, $\eta_i$ are coefficients which depend on the lattice scale
physics. The smallest of the three corresponds to the most
relevant operator in the dual formalism. The numerical result
gives that $\eta_3 = 11.3$ is the smallest coefficient, therefore
$V_{pl}$ is the most relevant operator in the dual formalism.

\section{Instabilities and resultant phases}

\subsection{Instability}

The algebraic phase has two dominant instabilites, which are
expected when the dimensionless coupling constant $K$ is very
large or very small.  If the pair tunnelling term in (\ref{ring})
is relevant, it breaks the symmetries listed in (\ref{symmetry})
and (\ref{symmetry2}) down to $\mathbb{Z}_2$ symmetries, which
then gaps out the gapless excitations of the algebraic phases. The
correlation functions between two pair-tunnelling terms
$\gamma\cos(2(\theta_{a} - \theta_b))$ vanishes within the
algebraic bond liquid phase, but two adjacent pair-tunnelling
terms can generate pair-bond operators such as $B_{2x} =
\cos(2\partial_x(\theta_x - \theta_y))$ and $B_{2ab,a} =
\cos(2\partial_a(\theta_a - \theta_b)) $. The correlation
functions between pair-bond operators take the form,

\beqn d = 2: \langle B_{2x}(0,0,0)B_{2x}(\tau,0,y)\rangle \sim
\frac{1}{|\tau^2 + y^2|^{4/(\pi^2K)}},\cr\cr d = 3: \langle
B_{2xy,x}(0,0,0,0)B_{2xy,x}(\tau,0,y,0)\rangle \cr\cr\sim
\frac{1}{|y^2 + \tau^2|^{4\eta^\prime/K}}. \cr \eeqn When the
pair-bond terms become relevant and grow in magnitude under coarse
graining, the $\gamma$ term itself will become important too, and
the system is expected to enter a $\mathbb{Z}_2$ phase, which we
discuss in the next subsection. The pair-bond terms are relevant
when $K > 2/\pi^2$ for $d = 2$, and $K > 2\eta^\prime$ for $d =
3$, so the $\mathbb{Z}_2$ phase will occupy the large $K$ region
of the phase diagram as shown in Fig. 6.

The vertex operators in the dual representation can drive another
instability when $K$ is small. Once relevant, these vertex
operators destabilize the algebraic phase, giving a Mott insulator
which will break lattice symmetries due to the Berry's phases. The
most relevant vertex operators in the dual representation depend
on the dimension and the average filling.  When $d = 2$ and
$\bar{n}_{a}$ are integers, the most relevant terms are $V_{x} =
\cos(2\pi\partial_x\varphi)$ and $V_y =
\cos(2\pi\partial_y\varphi)$; when $d = 2$ and $m$ is large
enough, the most relevant terms are parallel jump terms $V_{pl}$.
If the vertex operators are relevant, they will obviously break
the symmetry in (\ref{dualsymmetry}), since if we expand these
vertex operators, they would read \beqn V_{x} \sim
(\partial_x\varphi)^2, V_{y}\sim (\partial_y\varphi)^2,\cr\cr
V_{pl}\sim (\partial_x^2\varphi)^2 + (\partial_y^2\varphi)^2.\eeqn
These terms do not have the one dimensional symmetry any more.

\begin{figure}
\includegraphics[width=3.2 in]{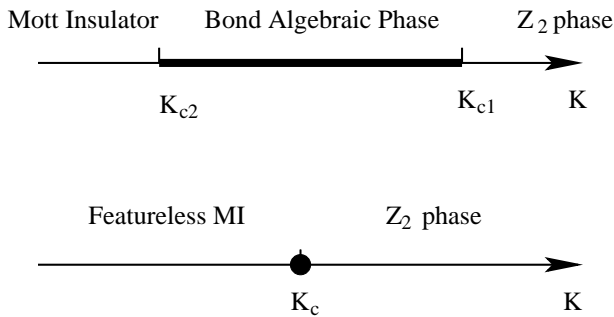}
\caption{The phase diagrams for the $p$-wave particles in both $d
= 2$ and $d = 3$ case. The top diagram is for the case with
$\bar{n}_x - \bar{n}_y = 2/m$, and $m > 1$ in $d = 2$; or $\bar{n}
\neq 3\mathbb{Z}$ in $d = 3$. In both cases nonzero Berry's phases
enter the vertex operators, determining the crystalline order in
the Mott insulator. The bottom diagram is for filling $\bar{n} =
d\mathbb{Z}$, in which the Berry's phase vanishes and the Mott
Insulator is featureless. Since the vertex operators are more
relevant when $\bar{n} = d\mathbb{Z}$, the algebraic bond liquid
does not exist as a stable phase.\label{phasedia}}
\end{figure}

When $d = 3$ and $\bar{n} = 3\mathbb{Z}$, the most relevant terms
are loop growing terms $G_{ab} = \cos(2\pi(\partial_a\varphi_b -
\partial_b\varphi_a))$, and when $d = 3$ and $\bar{n} \neq
3\mathbb{Z}$, the most relevant terms are again the $V_{pl}$
terms. Based on the correlation functions in (\ref{corre2d}) and
(\ref{corre3d}), the $V_{pl}$ terms are irrelevant when $K > 3/16$
for $d = 2$ and $K > 2/\eta_3$ for $d = 3$. The scaling dimensions
of $V_{1m}$ and $G_{13}$ operators are higher than the $V_{pl}$
terms as long as $m$ is big enough. A numerical evaluation of the
dimensionless numbers, which involves an integration over the full
Briullioun zone, gives $\eta_3 = 11.3$ and $\eta^\prime = 0.0950$.
Therefore there are finite parameter regions for the existence of
a stable algebraic bond liquid phase in both $ d = 2$ and $ d = 3$
(Fig. \ref{phasedia}): \beqn d = 2: 3/16 < K < 2/\pi^2, \cr\cr d =
3: 2/\eta_3 < K < 2\eta^\prime . \label{stable}\eeqn

However, when $\bar{n} = d \mathbb{Z}$ where the Berry's phase
terms vanish, lower order vertex operators are allowed which are
more relevant. In this case the bond algebraic liquid will not
exist as a stable phase (Fig. \ref{phasedia}).
One notable aspect of this theory is that the calculation of
the scaling dimensions depends on the behavior of the model on
the lattice scale.

\subsection{$\mathbb{Z}_2$ phase}

When the coupling $K$ is sufficiently large the $B_{2x}$ and
$B_{2ab,a}$ terms will be relevant, and the pair tunnelling term
in (\ref{ring}) will also grow large and will drive the system
into the $\mathbb{Z}_2$ phase. What is the nature of this phase?
First we notice that, even if the pair conversion term is
relevant, the symmetries in (\ref{symmetry2}) and (\ref{symmetry})
are still respected as long as $f_i$ and $g_i$ take only two
values, $0$ or $\pi$. In the following, we will first focus on the
$d = 2$ case, which is technically much simpler. One can directly
generalize to the $d = 3$ system.

Since $\gamma$ should be positive, when it becomes large it will
lock $\theta_x - \theta_y$ into two values, $\pm \pi/2$.  Because
$n_x + n_y$ is a static variable in the projected Hilbert space,
we can just focus on the variables $n_x - n_y$ and $\theta_x -
\theta_y$. Notice that if $n_x + n_y$ is even (odd), $n_x - n_y$
has to be even (odd) too.

\begin{figure}
\includegraphics[width=2.5 in]{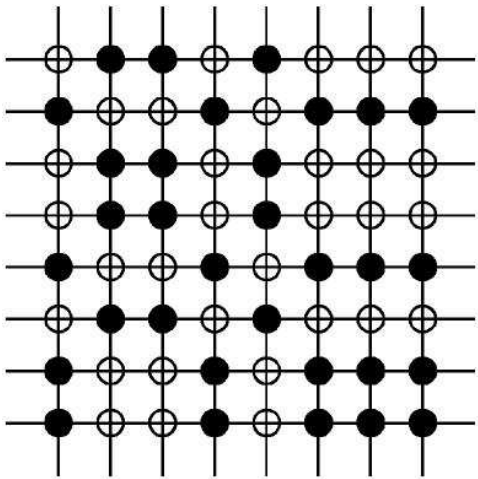}
\caption{One of the classical ground states of model (\ref{z2})
when $h = 0$. The ground states satisfy the condition that each
unit square on the lattice has either $(2+, 2-)$, or $(4+, 0-)$ or
$(0-, 4+)$ configurations. \label{z2fig}}
\end{figure}

Let us assume $n_x + n_y$ is even, for simplicity. Two
$\mathbb{Z}_2$ variables can be introduced to describe the system:
\beqn \sigma^z = i\exp(i(\theta_x - \theta_y)), \sigma^x =
\exp(i\pi (n_x - n_y)/2). \eeqn Using the formula \beqn e^Ae^B =
e^{[A,B]}e^Be^A \eeqn we can see $\sigma^x$ and $\sigma^z$
anticommute with one another, so they have the same algebra as
Pauli matrices.  In terms of these ``spin" variables, the original
model (\ref{ring}) can be effectively written as,
 \beqn H_{z2} =
\sum_i- h\sigma^x_i - \sum_\square
K\sigma^z_1\sigma^z_2\sigma^z_3\sigma^z_4 . \label{z2}\eeqn This
model was studied previously in reference [\cite{xu2004}], where
it was introduced to describe a phase transition in an array of
$p+ip$ superconducting grains. For $K > h$, it was shown that the
system is in a bond ordered phase, with the expectation value of
$\sigma^z_i\sigma^z_{i+\hat{x}}$ being independent of $y_i$. This
phase has a large degeneracy, $D$, which can be estimated by
counting the number of $\mathbb{Z}_2$ symmetries of
(\ref{symmetry2}).   The result is \beqn D = 2^{L_x +
L_y}.\label{z2dege}\eeqn The physical transformations within the
degenerate ground state manifold only transform $\theta_x -
\theta_y$. One of the classical ground states when $h = 0$ is
shown in Fig. \ref{z2fig} \cite{moore2004}. When $K < h$, there is
no bond order, and the ground state is not degenerate. The
transition between these two phases happens exactly at $K = h$.

In $ d = 3$ with $\gamma > 0$, the three pair tunnelling operators
are \beqn H_{v} = \gamma(\cos(2(\theta_{x} - \theta_{y}) +
\cos(2(\theta_{y} - \theta_{z}) + \cos(2(\theta_{z} - \theta_{x}))
\cr \eeqn These three terms cannot be independently minimized,
i.e. they are frustrated.  Minimizing the sum of all three terms
gives the following solutions, \beqn \theta_{x} - \theta_y = \pi
k_3/3, \theta_{y} - \theta_z = \pi k_1/3, \theta_{z} - \theta_x =
\pi k_2/3,\cr k_1 + k_2 + k_3 = 0, \eeqn where the $k_i$ are not
divisible by 3. The symmetry relating the manifold of solutions is
the same as in (\ref{symmetry}) except that $f_i + g_i$ only takes
two values, $0$ or $\pi$.

As in $d=2$, this $\mathbb{Z}_2$ phase exhibits bond order, with
spatially varying $B_{ab,a}$ and $B_{ab,b}$.  For instance, the
expectation value of $B_{xy,x} = \cos(\partial_x(\theta_x -
\theta_y))(x,y,z)$ is independent of $y$.  The expectation value
can be $\pm 1$ or $\pm 1/2$.  Once again, this $\mathbb{Z}_2$
phase has a large degeneracy. In the $\mathbb{Z}_2$ version of the
symmetry (\ref{symmetry}), the functions $f_i$ and $g_i$
correspond to physically different states in the ground state
manifold, but this is not the case for the function $\alpha$.  The
total degeneracy $D$ is obtained by counting all the physical
symmetry transformations, which is \beqn D\sim 2^{2(L_xL_y +
L_yL_z + L_zL_x + L_x + L_y+L_z)}\eeqn.

\begin{figure}
\includegraphics[width=2.8 in]{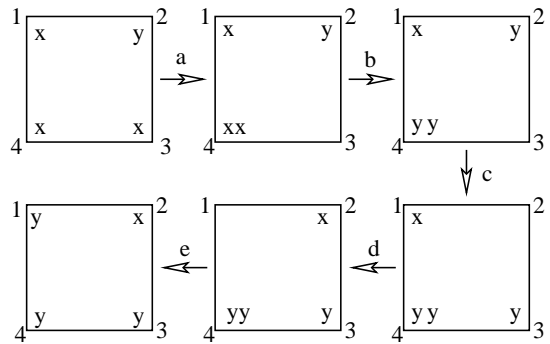}
\caption{The process in (\ref{n=1}) which involves four
intermediate states. Step ($a$), $p_x$ particle at site 3 hops to
site 4 (intermediate step); step ($b$), two $p_x$ particles at
site 4 can be transformed to two $p_y$ particles due to the pair
tunnelling term; step ($c$), $p_y$ particle at site $2$ hops to
site $3$; step ($d$), $p_x$ particle at site 1 hops to site 2;
step ($e$), one of the $p_y$ particles at site 4 hops to site
1).\label{fign=1}}
\end{figure}

When $\bar{n}=1$ the pair tunnelling term takes the system out of
the projected Hilbert space.  But acting in concert with hopping
processes it will generate terms such as  \beqn H_e =
\tilde{\gamma}\cos((\theta_x-\theta_y)_1 - (\theta_x-\theta_y)_2
\cr + (\theta_x-\theta_y)_3 + (\theta_x-\theta_y)_4).\label{n=1}
\eeqn Here 1,2,3 and 4 label the four corners of one unit square.
This process is shown in Fig. \ref{fign=1}. The correlation
function between two $H_e$ operators is zero in the algebraic
phase, but $H_e$ terms can generate terms which have nonzero
correlation functions along certain directions. If we check
carefully, we will find that the leading term generated is still
irrelevant enough for a stable algebraic liquid phase. Thus, with
$\bar{n}=1$ the algebraic boson liquid can exist as a stable phase.

What kind phase will $H_e$ drive the system into when it is
relevant?   Notice that equation (\ref{n=1}) has the same
$\mathbb{Z}_2$ symmetry as (\ref{z2}), with the particle number of
$n_x - n_y$ conserved mod 2 along every row and column. Thus the
order should again be bond order.  Indeed, if we take (\ref{n=1})
and the ring exchange term (\ref{ring}) together, the Hamiltonian
involving $\theta \equiv \theta_x - \theta_y$ reads, \beqn H =
-t\cos(\theta_1 - \theta_2 + \theta_3 - \theta_4) +
\tilde{\gamma}\cos(\theta_1 + \theta_2 + \theta_3 - \theta_4) \cr
+ \tilde{\gamma}\cos(- \theta_1 + \theta_2 + \theta_3 + \theta_4)
+ \tilde{\gamma}\cos( \theta_1 - \theta_2 + \theta_3 +
\theta_4)\cr + \tilde{\gamma}\cos( \theta_1 + \theta_2 - \theta_3
+ \theta_4) \cr. \eeqn One of the minima of this Hamiltonian is
$\theta_i = \pi / 2$.  All of the other minima can be obtained by
transforming $\theta_i \rightarrow \theta_i + \pi$ on each row and
column. The symmetry and the ground state are thus the same as
(\ref{z2}).  The resulting phase is precisely the $\mathbb{Z}_2$
phase found before.

\subsection{Mott Insulator phase}

For small enough coupling $K$, when the vertex operators in the
dual formalism are relevant, the system is driven into the Mott
insulator phase. In the integer filling case with $\bar{n} =
d\mathbb{Z}$, the Mott insulators are featureless and there is a
unique ground state with no broken symmetries. With fractional
filling, flavor-density ordered patterns which breaks space
symmetries are expected. Again, let us first focus on $d = 2$
case, and then generalize the results to $d = 3$.

In the 2 dimensional Mott Insulator phase, because the average
filling of each flavor of bosons could be fractional,
crystalline order is expected. However, the crystalline order is
only in the $n_x - n_y$ channel, since $n_x + n_y$ is fixed to be
$\bar{n}$. The dual action in the case with $\bar{n} =
2\mathbb{Z}$ and $\bar{n}_x - \bar{n}_y = 2/m$ reads,

\begin{eqnarray}
L_{loop} = \frac{1}{2K}(\partial_\tau\varphi)^2 +
\frac{1}{2K}(\partial_x\partial_y\varphi)^2\cr\cr -
\tilde{\gamma}\cos(2m\pi\partial_x\varphi) -
\tilde{\gamma}\cos(2m\pi\partial_y\varphi) + \cdots \label{loop2}
\end{eqnarray}

The ellipses include other vertex operators, for instance the
$V_{pl}$ terms, however, these terms do not determine the ordered
pattern. The double vertex operators induce degenerate ground
state once they are relevant. The most natural phases have stripe
order in the $n_x - n_y$ density or a  plaquette order.   The
order parameters can be represented in terms of low energy
variables:

\begin{eqnarray}
(n_x - n_y - \bar{n}_x + \bar{n}_y)e^{2\pi ix/m} \sim
\sin(2\pi\partial_y\varphi),\cr (n_x - n_y - \bar{n}_x +
\bar{n}_y)e^{2\pi iy/m}\sim \sin(2\pi\partial_x\varphi);\cr\cr
\cos(\partial_x\partial_y(\theta_x - \theta_y))e^{2\pi ix/m}\sim
\cos(2\pi\partial_y\varphi), \cr
\cos(\partial_x\partial_y(\theta_x - \theta_y))e^{2\pi iy/m}\sim
\cos(2\pi\partial_x\varphi).\label{stripe}
\end{eqnarray}

\begin{figure}
\includegraphics[width=2.1 in]{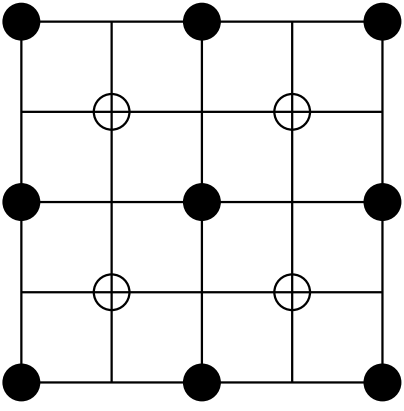}
\caption{The stripe order of $n_x - n_y$ in the case of $\bar{n}$
is odd integer and $\bar{n}_x = \bar{n}_y$. The full circles
represent sites with $n_x - n_y > 0$, and the empty circles
represent sites with $n_x - n_y < 0$. \label{00stripe1}}
\end{figure}

This representation can be obtained from either symmetry arguments
or from renormalization group flow of the lattice theory. If the
vertex operators in (\ref{loop2}) are relevant, the stripe order
parameters in (\ref{stripe}) take nonzero values. If
$\tilde{\gamma}$ in (\ref{loop2}) is positive, the plaquette
stripe order is favored; if $\tilde{\gamma}$ is negative, the $n_x
- n_y$ density stripe order is favored. If $\bar{n}$ is odd
integer and $\bar{n}_x = \bar{n}_y$, the dual action and the low
energy representation of order parameter are similar to the
equations (\ref{loop2}) and (\ref{stripe}) with $m = 2$. The $n_x
- n_y$ density stripe order with odd integer $\bar{n}$ is depicted
in Fig. \ref{00stripe1}. Both density stripe order and plaquette
stripe order have been found numerically in a spin-1/2 model with
a similar 1 dimensional gauge symmetry
\cite{sandvik2003,sandvik2004}. Notice that the crystalline
pattern predicted from the field theory is only valid when the
correlation length is much longer than the lattice constant, i.e.
one is close to the phase transition between the Mott and
algebraic liquid phases.

In $d = 3$, in the case of $\bar{n} \neq 3\mathbb{Z}$ and
$\bar{n}_x = \bar{n}_y = \bar{n}_z = \bar{n}/3$, the dual action
including the vertex operators reads,
\begin{eqnarray}
L_{loop} = \cr \cr\sum_a\frac{1}{2K}(\partial_\tau\varphi_a)^2 +
\sum_{a \neq b, b \neq c, c \neq
a}\frac{1}{2K}(\partial_a(\partial_b\varphi_c -
\partial_c\varphi_b))^2\cr - \sum_{a\neq b}g\cos(6 \pi(\partial_a\varphi_b - \partial_b\varphi_a)) + \cdots
\label{loop} \cr \end{eqnarray} The triple vertex operators gives
rise to degenerate ground state when they are relevant. Again,
because $n_x + n_y + n_z$ is fixed at $\bar{n}$, the density
modulation can only happen in the $n_a - n_b$ variables. The most
natural state is the stripe order, and the order parameter can be
represented in terms of low energy variables as,

\begin{eqnarray}
n_{x} - n_{y} \sim \exp(2\pi ix/3)\sin(2\pi(\partial_y\varphi_z -
\partial_z\varphi_y)) \cr - \exp(2\pi iy/3)\sin(2\pi(\partial_z\varphi_x -
\partial_x\varphi_z)),\cr\cr n_{y} - n_{z} \sim \exp(2\pi iy/3)\sin(2\pi(\partial_z\varphi_x -
\partial_x\varphi_z)) \cr - \exp(2\pi iz/3) \sin(2\pi(\partial_x\varphi_y -
\partial_y\varphi_x)),\cr\cr n_{z} - n_x \sim \exp(2\pi iz/3)\sin(2\pi(\partial_x\varphi_y -
\partial_y\varphi_x)) \cr - \exp(2\pi ix/3)\sin(2\pi(\partial_y\varphi_z -
\partial_z\varphi_y)) .\label{stripe3}
\end{eqnarray}

The plaquette stripe order parameters can be obtained in a similar
way. These order parameters take nonzero expectation value once
the vertex operators in (\ref{loop}) are relevant. Because the
order parameters in (\ref{stripe3}) are not invariant under the
space symmetry group, the ground states will have spontaneously
broken symmetries.

A small off-site density-density interaction between bosons can in
principle exist, e.g. the $\delta H$ and $\delta H^\prime$ terms
discussed in the second section. In the bond liquid phase, a term
of this form renormalizes the field theory parameter $K$, but it
does not introduce any new vertex operators. Thus a perturbation
of off-site density interaction does not destabilize the liquid
phase.

\subsection{Other instabilities}

Throughout the paper, we have assumed that the hopping amplitudes,
$t_{a,b,c}$ in (\ref{atomhop}), are only non-zero when $a = b =c$.
In a real system, $t_{a,b,c}$ will also be non-zero when this
condition is not satisfied, but are expected to be very small.
Once non-zero, these hopping processes will lead to  new
instabilities of the bond algebraic liquid phase.   For instance,
at the order of $t_{x,x,x}t_{y,y,x}/U$, bond operator $B_{xy,x} =
\cos(\partial_x(\theta_x - \theta_y))$ is generated.  Once present
this term will be relevant in the bond algebraic liquid phase. The
system will be driven into a phase where $e^{i(\theta_x -
\theta_y)} = \pm i$, the ground state is two fold degenerate. But
so long as these other hopping amplitudes are very small, this
instability will only occur at very low temperatures, and at
higher temperatures the system will behave as if it were in the
algebraic liquid phase. For instance, at small temperature the
system can be in the quantum critical region controlled by
critical point at which these perpendicular hopping terms are
tuned to zero, this quantum critical point is described by the
Gaussian actions (\ref{gaussian}).

\section{other phases}

\subsection{Doping with $s-$wave particles}

The results in the previous sections have been obtained under the
assumption that all the particles are in $p-$wave states.  But
what behavior is expected if some of the particles are in the
$s-$wave ground state? Let us take $ d = 2$ as an example.  Let us
suppose that the total lattice filling is still an integer
$\bar{n}$. For strong repulsive interactions which preclude
superfluidity, we again project into the constrained Hilbert space
with $\sum_a n_a + n_s = \bar{n}$ on every site of the lattice.
Here $n_s$ is the density of  $s-$wave particles, and we will
denote the conjugate phase field as $\theta_s$.

Since the $s-$wave particles can readily hop in all directions,
the low energy effective hopping within the projected Hilbert
space will allow a term of the form (assuming $d=2$ for
simplicity): \beqn H_{h} = -t_1\cos(\partial_x\theta_s -
\partial_x\theta_x) -t_1\cos(\partial_y\theta_s -
\partial_y\theta_y) .
\label{dopes}\eeqn  The presence of the $s-$wave particles can
thus strongly modify the dynamics of the $p-$wave particles.
Moreover, in addition to the allowed pair tunneling between the
$p_x$ and $p_y$ flavor particles, it will also be possible to
exchange $p-$wave and $s-$wave particles. Although a single
particle tunnelling is still forbidden by parity symmetry, pair
tunnelling between $s-$wave and $p-$wave particles will be
allowed. Upon including these processes, the full Hubbard-like
Hamiltonian in the rotor representation will take the form, \beqn
H_{r} = -t\cos(\partial_x\theta_s - \partial_x\theta_x)
-t\cos(\partial_y\theta_s - \partial_y\theta_y) \cr \sum_{a =
x}^yu_p(n_a - \bar{n}_p/2)^2 + u_s(n_s - \bar{n}_s)^2  \cr - \sum_{a
= x}^y g_a \cos(2(\theta_s - \theta_a)) - g_{xy} \cos(2(\theta_x -
\theta_y)) \eeqn Here $\bar{n}_s$ and $\bar{n}_p$ are the total
fillings for the $s-$wave and $p-$wave particles, with $\bar{n}_s
+ \bar{n}_p = \bar{n}$. We are implicitly assuming that $\bar{n}
>1$.

Consider first the situation in which the $s-$wave particles are
at integer filling and have a charge gap, which will be the case
for large enough $u_s$. The above hopping term will then take the
system out of the low energy sector, but can generate an effective
ring exchange term, \beqn H_{ring} =
-t\cos(\partial_x\partial_y(\theta_x - \theta_y) ) + \cdots .
\eeqn This term is the same as the ring exchange term in
(\ref{ring}).   Therefore when the  $s-$wave particles are gapped,
the system behaves similarly to the undoped case, as expected on
physical grounds.

To address the more interesting situation in which
$\bar{n}_s,\bar{n}_p$ are non-integer, it will be helpful to first
consider the symmetries of the above Hamiltonian in the absence of
the pair tunneling terms, $g_a = g_{xy} =0$.  In this case the
model is invariant under, \beqn \theta_x \rightarrow \theta_x +
f_1(y) + g_1(x)+ \alpha(x,y), \cr \theta_y \rightarrow \theta_y +
f_2(x) + g_2(y) + \alpha(x,y), \cr \theta_s \rightarrow \theta_s +
g_1(x) + g_2(y) + \alpha(x,y).\eeqn

We now search for a stable algebraic phase by first setting $g_a =
g_{xy} =0$, and then expanding the cosine terms in $H_r$ to obtain
the following quadratic Lagrangian, \beqn L_{d} = \sum_{\mu =
x,y,s} \frac{K}{2}(\partial_\tau\theta_\mu)^2 + \sum_{a =
x}^y\frac{K}{2}(\partial_a(\theta_s - \theta_a))^2
.\label{dopeaction2}\eeqn Upon diagonalization, one obtains 3
modes, which at small $\vec{k}$ are given by, \beqn \omega_0^2 =
0,\cr\cr \omega_1^2 \sim k_x^2 + k_y^2 - \sqrt{k_x^4 - k_x^2k_y^2
+ k_y^4 } \sim \frac{k_x^2k_y^2}{k^2},\cr\cr \omega_2^2 \sim k_x^2
+ k_y^2 + \sqrt{k_x^4 - k_x^2k_y^2 + k_y^4 } \sim
k^2.\label{dopemode2}\eeqn Again $\omega_0$ corresponds to the
unphysical gauge mode resulting from the hard constraint $\sum_a
n_a + n_s = \bar{n}$.  The $\omega_2$ mode corresponds to the
gapless Goldstone mode resulting from the ``condensation" of
$\theta_s$.   The $\omega_1$ mode vanishes along the coordinate
axis in momentum space, and it gives rise to quasi-1d physics.

The relevance of the pair tunnelling terms can be estimated by
considering their scaling dimension in the Gaussian phase action
(\ref{dopeaction2}).   Based on the above symmetries of the
Gaussian theory, it is concluded that the operators
$\cos(2(\theta_s - \theta_a))$ can only correlate in one
dimension.   Moreover, a calculation shows that it has a power law
correlation function, due to the contribution from $\omega_1$ in
(\ref{dopemode2}).  The parallel jump term generated by
$\cos(2(\theta_a - \theta_b))$ is not as relevant as
$\cos(2(\theta_s - \theta_a))$.  When $\cos(2(\theta_s -
\theta_a))$ is irrelevant, the system is in another type of
algebraic liquid phase different from the bond liquid phase, which
is described by the Gaussian action (\ref{dopeaction2}). The
operators which have power law correlations are the tunneling
operators $\cos(2(\theta_s - \theta_a))$ instead of bond
operators.

\subsection{1d liquid phase and superfluid phase}

So far everything we have considered is under the constraint
$\sum_a n_a = \bar{n}$ on every site of the lattice. What will
happen if we soften this constraint?  Softening the constraint
will lower the gap for the charged excitations, and eventually
superfluid order can develop which breaks the global $U(1)$
symmetry of the phase $\sum_a \theta_a$.   This superfluid phase
has been discussed in reference \cite{girvin2005}.  But as we now
show it is also possible to have a stable quasi one-dimensional
liquid phase.

Let us take the $d = 2$ system as an illustrative example. The
original Hamiltonian before projection is (\ref{atomfull}) . We notice that, with
the pair tunnelling term, the symmetry of the system is the
$\mathbb{Z}_2$ version of the 1d symmetry, and there is moreover a
global 2 dimensional $U(1)$ symmetry $\theta_x \rightarrow
\theta_x + \theta_0$, $\theta_y \rightarrow \theta_y + \theta_0$,
for a spatially independent $\theta_0$. In the case with small
interaction $U$, upon passing to the rotor representation, and
expanding the cosine hopping terms, the action associated with the
Hamiltonian in (\ref{atomfull}) becomes,
 \beqn L =
\frac{1}{2K}  \sum_{a=x,y} [ (\partial_\tau \theta_a)^2 +
(\partial_a \theta_a)^2 ]  \cr - \gamma\cos(2\theta_x -
2\theta_y). \eeqn

When $\gamma=0$ this Lagrangian describes a type of
one-dimensional liquid phase, which can be visualized as a set of
decoupled 1 dimensional Luttinger liquids on every row and column
of the 2 dimensional square lattice. To examine the stability of
these Luttinger liquids one must check to see when the vortex
tunneling events in the dual representation are
irrelevant\cite{kane1992}. If we assume $\bar{n}$ is odd, the
average filling for each flavor is half-integer, and the single
vortex operator in the dual theory has a nonzero Berry's phase.
Double strength vortices are irrelevant provided\cite{kane1992} $
K < 2\pi$.   For $K>2\pi$ the proliferation of these vortices will
destabilize the quasi 1d liquid phase and drive the system into a
Mott insulator with broken translational symmetry.

Stability of this quasi-1d liquid phase also requires the
irrelevance of $\gamma$.  First we note that the two-point
correlation function of the operator $\cos(2\theta_x - 2
\theta_y)$, when evaluated in the quadratic theory, vanishes
unless the operators are on the same site. We can thus focus
exclusively on the time dependence of this correlator at (any) one
given point on the 2d lattice, say $x=y=0$. This site lies on only
two of the Luttinger liquids, one running in the $x-$direction and
the other in the $y-$direction. This problem can then be viewed as
a zero dimensional point contact between two Luttinger liquids,
which has been studied in detail\cite{kane1992}. Following this
work we find that $\gamma$ is irrelevant when $ K > \pi$. We thus
conclude that when $\bar{n}$ is odd, there will be a range of
parameters, $ \pi < K < 2\pi$, within which the quasi-1d liquid
exists as a stable phase.

What if $\gamma$ is relevant? In this case we should minimize the
cosine potential, which will lock the two fields $ \theta_{ix} =
\theta_{iy} + \eta_i\pi/2$. $\eta_i = \pm 1$, and can be
represented as $\eta_i = (-1)^{f(i_x) + g(i_y)}$. $f$ and $g$ are
both integer valued functions. In the simplest case wherein $f$
and $g$ are both constants, then the quadratic part of the
Lagrangian when re-expressed in terms of a new field, $\theta =
\theta_x + \theta_y $, will be simply $L \sim \sum_{\mu=x,y,\tau}
(\partial_\mu \theta)^2$. This is the Lagrangian of a conventional
2d superfluid, found in previous work \cite{girvin2005}.  We thus
conclude that the $\gamma$ term tends to drive the system from the
quasi-1d liquid phase into a 2 dimensional superfluid phase.

\section{experimental implications}

Here we  discuss how one might try to detect experimentally the
presence of the algebraic bond liquid phase. We will consider
three different types of correlation functions which can in
principle be measured in cold atom experiments. First, we will
consider the particle density-density correlator, which can in
principle be extracted via light scattering experiments. Secondly,
we will consider the momentum distribution function, $n_k =
\langle \psi^\dagger_k \psi_k \rangle$. In cold atom experiments
this quantity can be extracted by releasing the atoms from the
trap, and at some later time measuring their spatial distribution
function in real space,  $\tilde{n}(R)$, with $R$ the radial
coordinate measured from the center of the trap. At time $t$ after
releasing the atoms, these two quantities are related as,
$n_{k=mR/t} = \tilde{n}(R)$. Finally, we consider the momentum
density - momentum density two point function, $\langle n_k
n_{k^\prime} \rangle$, which can be measured by spatial noise
correlations \cite{demler2004,folling2005}.

\subsection{Particle density-density correlation function}

The particle density-density correlation function $\langle
\rho(0)\rho(r)\rangle$ can be used to distinguish between various
phases.  For example, the crystalline Mott insulator phases
manifest themselves as Bragg peaks, whereas quasi-1d liquid phase
results in a power law behavior. The Bragg scattering techniques
have been applied to atoms trapped in an optical lattice
\cite{folling2005}, and the periodicity of optical lattice was
extracted. The particle density-density correlation function
depends on the form factor of $p-$wave particles, and the flavor
density-density correlation function (for instance $\langle n_x(i)
n_y(j)\rangle$). In the phase of primary interest - the algebraic
liquid phase - the correlator between total flavor density
operators provides no useful information since the total particle
density on each site is fixed at $\bar{n}$. Thus, the correlation
functions between the total density operator $n_x + n_y$ will be
quite trivial. However, since $p_x$ and $p_y$ particles have
different form factors, the particle density-density correlation
function $\langle \rho(0),\rho(r)\rangle$ contains nontrivial
information about this algebraic phase.

We first briefly consider here the behavior of the correlator of
the flavor-density operator, which is non-trivial in the
algrebraic liquid phase. From the $d=2$ duality we have, \beqn n_x
- \bar{n}_x =
\partial_x\partial_y\phi , \cr n_y - \bar{n}_y = -
\partial_x\partial_y\phi . \eeqn
In terms of the coarse grained field $\varphi$, the representation
of $n_x$ and $n_y$ are \beqn n_x \sim \bar{n}_x +
\partial_x\partial_y\varphi \cr\cr +
c(e^{2\pi ix/m}\sin(2\pi\partial_y\varphi) + e^{2\pi
iy/m}\sin(2\pi\partial_x\varphi)), \cr\cr  n_y \sim \bar{n}_y -
\partial_x\partial_y\varphi \cr\cr - c(e^{2\pi ix/m}\sin(2\pi\partial_y\varphi)
+ e^{2\pi iy/m}\sin(2\pi\partial_x\varphi)).\label{corre}\cr\eeqn
The correlation functions can then be readily computed giving,
\beqn \langle (n_x(0,0,0) - \bar{n}_x)(n_x(x,0,\tau) -
\bar{n}_x)\rangle \cr \sim e^{2\pi ix/m}\frac{1}{|x^2 +
\tau^2|^{4K}} + a\frac{1}{|x^2 + \tau^2|},\cr\cr\cr \langle
(n_x(0,0,0) - \bar{n}_x)(n_x(0,y,\tau) - \bar{n}_x)\rangle \cr
\sim e^{2\pi iy/m}\frac{1}{|y^2 + \tau^2|^{4K}} + a\frac{1}{|y^2 +
\tau^2|},\cr\cr\cr \langle (n_x(0,0,0) - \bar{n}/2)(n_x(x,y,0) -
\bar{n}/2)\rangle\sim \frac{1}{x^2y^2}.
 \label{corre2}\eeqn

When the vector separating the two $n_x$ operators is parallel to
one of axes, $x$ or $y$,  the correlation function has two
contributions. The first comes from the sine term, the second from
the derivative term in (\ref{corre}).

When the vector separating the two $n_x$ operators is not along an
axis, only the derivative term in (\ref{corre}) contributes.
Notice that $n_x$ not only correlates in the $x$ direction, but
also in the $y$ direction. This is due to the fact that,the ring
exchange term in (\ref{ring}) has no asymmetry between $x$ and
$y$. However, in the correlation function (\ref{corre2}) the two
dimensional rotational symmetry is not restored even when the
scale is much larger than the lattice constant, which is quite
different from other algebraic liquid phases.

The particle density correlation functions are supposed to be
measurable with light scattering experiment. The result is usually
written in momentum space $ \langle \rho_k,\rho_{-k}\rangle $,
which contains information of both the form factor of $p-$level
state wave functions, and the particle density correlation
functions. In the case of the bond-algebraic phase, the result at
small momentum is \beqn \langle\rho_k,\rho_{-k}\rangle \cr\cr \sim
\sum_{i,j}\sum_{a,b = x}^y \langle n_{ia},n_{jb} \rangle\times
\int d^2xd^2y e^{i\vec{k}\cdot(\vec{x} -
\vec{y})}|\phi_{ia}(\vec{x})|^2|\phi_{jb}(\vec{y})|^2 \cr\cr \sim
F(k_x,k_y)\times (k_x^2 - k_y^2)^2  +
\sum_ic_1\delta(\vec{k}-\vec{G}_i).\cr\eeqn
Here $\vec{G}_i$ are the
vectors in the reciprocal lattice of the optical trap potential, and $F(k_x, k_y)$ is a function
of $k_x$ and $k_y$, \beqn F(k_x,k_y) = |k_xk_y| + \sum_{a =
x}^yc(k_a - 2\pi/m)^{8K - 1}.\cr \label{correfunc}\eeqn The
$\phi_{ia}(\vec{x})$ is the wave function of the atoms at state
$p_a$ on site $i$.

In the case of Mott Insulator phase with a crystalline pattern of
flavor difference $n_x - n_y$, the result is \beqn
\langle\rho_k,\rho_{-k}\rangle \sim \sum_i \delta(\vec{k} -
\vec{G^\prime_i})(k_x^2 - k_y^2)^2 +
\sum_ic_2\delta(\vec{k}-\vec{G}_i). \cr \eeqn
Here $\vec{G}^\prime_i$
are the crystalline order wave vectors. Therefore we conclude that
using light scattering experiment, one can detect the bond
algebraic phase.

\subsection{Momentum distribution function}

The momentum distribution function has been used to detect both
the Bose-Einstein condensate, being characterized by a sharp peak
at $k = 0$, and the Mott insulator where there is no sharp
peak\cite{greiner2002}.  We can calculate the momentum
distribution function of various phases as follows. Focussing for
simplicity on 2 dimensional systems, the momentum distribution for
the superfluid phase of $s-$wave particles is \beqn n(\vec{k})/N =
1/N \int d^2r_1d^2r_2 \langle\psi^\dagger(\vec{r}_1)
\psi(\vec{r}_2) \rangle e^{i\vec{k}(\vec{r}_1 - \vec{r}_2)} \cr\cr
\sim 1/N \int d^2r_1d^2r_2 \exp(i\vec{k}\cdot(\vec{r}_1 -
\vec{r}_2)) \cr\cr \times \sum_{i,j} \langle \phi_{si}(\vec{r}_1)
d^{(s)\dagger}_i \phi_{sj}(\vec{r}_2) d^{(s)}_j \rangle \cr\cr =
1/N \sum_{i,j} e^{i\vec{k}\cdot(\vec{R}_i -
\vec{R}_j)}|\phi_s(k)|^2 \langle d^{(s)\dagger}_id^{(s)}_j \rangle
\cr\cr \sim \sum_i |\phi_s(\vec{k})|^2\delta(\vec{k} -
\vec{G}_i)\eeqn Here $\vec{G}$ is the vector in the reciprocal
lattice of the optical lattice, $\phi_s(\vec{k})$ is the Fourier
transformation of the real space $\phi_s(r)$, which roughly
behaves as $\phi_s(\vec{k}) \sim const.$ at small $k$. The
momentum distribution function of the superfluid phase of $p-$wave
particles has been calculated in \cite{girvin2005}. The result is
complicated by the fact that in this phase there is large set of
$\mathbb{Z}_2$ degeneracies, as shown in (\ref{z2dege}).   The
result of the calculation depends on which of the degenerate
states the system is in.

For a Mott Insulator phase of $s-$wave particles, the momentum
distribution is \beqn n(\vec{k})/N \sim |\phi_s(\vec{k})|^2. \eeqn
For a Mott insulator of $p-$wave particles, one finds, \beqn
n(\vec{k})/N \sim (a |\phi_x(\vec{k})|^2 + b |\phi_y(\vec{k})|^2).
\eeqn Here, the magnitudes of $a$ and $b$ depend on the relative
abundance of $p_x$ and $p_y$ particles. At small momentum,
$\phi_x(\vec{k}) \sim k_x$, $\phi_x(\vec{k}) \sim k_y $.

Although the algebraic bond liquid and the Mott Insulator phases
are qualitatively different, the momentum distribution functions
are the same, \beqn n(\vec{k})/N \sim
(|\phi_{x}(\vec{k})|^2\bar{n}_x + |\phi_{y}(\vec{k})|^2\bar{n}_y).
\label{momentumdis}\eeqn The reason for this is that the single
particle correlation function is short-ranged in both the Mott
insulator and the algebraic bond liquid nature, behaving in the
latter case as $\langle d^{(x)\dagger}_id^{(x)}_j\rangle \sim
\delta_{ij}$. Thus the momentum distribution function cannot
distinguish between these two insulating phases.

\subsection{Momentum density - momentum density correlator}

Recently it has been proposed that the momentum density - momentum
density correlation function \beqn \langle n(R_1)n(R_2)\rangle
\sim \langle n_{Q_1}n_{Q_2} \rangle  , \label{noise}\eeqn with
$\vec{Q}_i = m\vec{R}_i/t$ can be extracted in cold atom systems
by measuring spatial noise correlations \cite{demler2004}.
Remarkably, this technique has been successfully implemented to
detect crystalline density order\cite{folling2005}.

In the superfluid phase of $p-$wave particles, this correlator has
been calculated in \cite{girvin2005}, and the result again depends
on which of the $\mathbb{Z}_2$ degenerate  ground states the
system is in. As we now show, this correlator can in principle be
used to distinguish between the algebraic bond liquid and the Mott
insulator phases.  The calculation is easier to carry out for the
normal ordered operator, \beqn \langle
\psi^\dagger_k\psi^\dagger_{k^\prime}\psi_k\psi_{k^\prime}
\rangle/N^2 = \langle n_kn_{k^\prime}\rangle/N^2 + \langle n_k
\rangle\delta(\vec{k} - \vec{k}^\prime)/N^2 ,  \cr \eeqn
and $\langle
n_k\rangle / N$ is given in (\ref{momentumdis}). The calculation
of the correlation function in the algebraic bond liquid phase is
lengthy but straight forward.  At small $k$ we find, \beqn \langle
n_{\vec{k}}n_{\vec{k^\prime}} \rangle / N^2 \cr\cr \sim
a(\bar{n}_xk_x^2 + \bar{n}_yk_y^2)(\bar{n}_xk^{\prime 2}_x +
\bar{n}_yk^{\prime 2}_y) \cr\cr + \sum_i b(k_xk^\prime_x\bar{n}_x
+ k_yk_y^\prime\bar{n}_y)^2\delta(\vec{k}-\vec{k}^\prime +
\vec{G}_i) \cr\cr + c_1(k_x^2 - k_y^2)(k_x^{\prime 2} -
k_y^{\prime 2})/N \cr\cr + \alpha (k_xk_x^\prime -
k_yk_y^\prime)^2F(k_x - k_x^\prime, k_y - k_y^\prime)/N \cr\cr -
\gamma k^2\delta(\vec{k} - \vec{k}^\prime)/N + \cdots\cdots \cr
\label{densitycorre} \eeqn Here $N$ is the total number of sites
in the system, and the ellipses include all the terms smaller than
$O(1/N)$. $F$ is the function defined in (\ref{correfunc}).
Vectors in the reciprocal lattice of the optical lattice are
denoted as $\vec{G}$. The algebraic correlation function between
flavor density give rise to the third and forth line in equation
(\ref{densitycorre}).

In the stripe ordered Mott Insulator phase defined in
(\ref{stripe}), the momentum density - momentum density
correlation function reads, \beqn \langle
n_{\vec{k}}n_{\vec{k^\prime}} \rangle/N^2 \cr\cr \sim
a(\bar{n}_xk_x^2 + \bar{n}_yk_y^2)(\bar{n}_xk^{\prime 2}_x +
\bar{n}_yk^{\prime 2}_y) \cr\cr + \sum_i b(k_xk^\prime_x\bar{n}_x
+ k_yk_y^\prime\bar{n}_y)^2\delta(\vec{k}-\vec{k}^\prime +
\vec{G}_i) \cr\cr + \sum_ic_2(k_xk_x^\prime -
k_yk_y^\prime)^2\delta(\vec{k} - \vec{k}^\prime +
\vec{G}^\prime_i) \cr\cr - \gamma k^2\delta(\vec{k} -
\vec{k}^\prime)/N + \cdots , \cr \eeqn where $\vec{G}^\prime$ is
the ordering wave vector of the stripe order. Therefore, this
correlation function can be used to distinguish between the Mott
insulator and the algebraic bond liquid phase. At the leading
order, the two phases differ by the ordering wave vector
$\vec{G}^\prime$.  There are also qualitative differences in the
$O(1/N)$ terms at small $k$.

A more complicated correlation function which can in principle be
used to directly measure the power law correlations in the
algebraic bond liquid phase is $\langle n^2_kn^2_{k^\prime}
\rangle$. We find, \beqn \langle n_k^2n_{k^\prime}^2 \rangle/N^4
\sim O(1) + \cdots + \alpha (1/N^3) k_x^2k_y^2k^{\prime
2}_xk^{\prime 2}_y\cr\cr\times (|k_x - k_x^\prime|^{2/(\pi^2K) -
1} + |k_y - k_y^\prime|^{2/(\pi^2K) - 1} \cr\cr + |k_x +
k_x^\prime|^{2/(\pi^2K) - 1} + |k_y + k_y^\prime|^{2/(\pi^2K) -
1}). \eeqn The coefficient $K$ is defined in the Gaussian action
(\ref{gaussian}), and it directly encodes the power law decays in
the algebraic bond liquid phase.

\section{conclusions and extensions}

In this work, a novel bosonic liquid phase has been predicted and explored
in a system with
strongly correlated bosons in different orbital levels
of the wells of an optical lattice. In this algebraic bose liquid
phase the correlation functions between two bond operators fall off
as a power of the spatial separation.
The reason for the existence of this new phase is
the spatial anisotropy of the orbital wave functions: the $p$-level
states extend and hop preferentially in one direction. Due to this orbital
anisotropy, such a system in either $d=2,3$ can
behave similarly to a one
dimensional system. This effect has been called ``dimensional
reduction" \cite{xu2005,nussinov2006}. From the theoretical point
of view, the phases studied in this work are new bosonic liquid
phases in dimensions higher than one, which broadens our
understanding of gapless ``insulating" bosonic phases.
Experimentally, it is an open challenge to see if such algebraic phases can
be achieved in the cold atoms context.

A possible extension of this work is to consider a fermionic
version of the model studied herein. Fermionic atoms such as
$^{40}\mathrm{K}$ and $^{6}\mathrm{Li}$ trapped in an optical
lattice have been studied extensively recently, both theoretically
and experimentally. Various behaviors have been predicted - and
some observed -  in such fermionic atomic systems. For instance,
the normal fermi surface \cite{kohl2005}, the condensate of
molecules $^{40}\mathrm{K}_2$ and $^{6}\mathrm{Li}_2$
\cite{greiner2002,Jochim2003}, the fermionic superfluid state with
and without imbalanced fermion populations
\cite{ketterle2006,ketterle2005} and the crossover between
molecular condensate and the BCS like cooper pair condensate (for
instance, see \cite{ohashi2002}).

If fermionic atoms are pumped to $p-$level states, when
interaction between atoms keeps the filling on each well close to
$\bar{n}$, we can again effectively study the low energy projected
Hilbert space with the constraint. The ring exchange term similar
to (\ref{ringd}) is generated at low energy. New fermionic phase
different from all the previous studies is expected in this case,
characterized by novel non-fermi liquid behaviors.

The authors thank Eugene Demler for useful discussions.

\bibliography{p-wave}
\end{document}